\documentclass[12pt]{article}
\usepackage{epsfig}
\usepackage{amsfonts,amsthm}
\usepackage{amsmath,amssymb}
\usepackage{hyperref}

\newcommand{\be}{\begin{equation}}
\newcommand{\ee}{\end{equation}}
\newcommand{\bea}{\begin{eqnarray}}
\newcommand{\eea}{\end{eqnarray}}

\renewcommand{\Re}{\mathrm{Re }}
\renewcommand{\Im}{\mathrm{Im }}

\newcommand{\doublet}[2]{ \left( \begin{array}{c}#1 \\ #2 \end{array}\right) }

\newcommand{\lr}[1]{ \langle #1 \rangle}

\providecommand{\calV}{\mathcal{V}}

\providecommand{\RR}{{\mathbb{R}}}
\providecommand{\vr}{\vec{r}}
\providecommand{\tK}{\tilde{K}}
\providecommand{\ts}{\tilde{s}}
\providecommand{\id}{\boldsymbol{1}}
\def\Tr{\mathrm{Tr}}
\providecommand{\tp}{\mss{\mathsf{T}}}
\providecommand{\diag}{\mathrm{diag}}
\providecommand{\eq}[1]{\begin{equation} #1 \end{equation}}
\providecommand{\eqarr}[1]{\begin{eqnarray} #1 \end{eqnarray}}
\providecommand{\mss}[1]{\mbox{\scriptsize $#1$}}
\providecommand{\mfn}[1]{\mbox{\footnotesize $#1$}}
\providecommand{\ms}[1]{\mbox{\small $#1$}}
\providecommand{\ml}[1]{\mbox{\large $#1$}}
\providecommand{\bs}[1]{\boldsymbol{#1}}
\providecommand{\hs}[1]{\hspace{#1}}

\def\lsim{\mathrel{\rlap{\lower4pt\hbox{\hskip1pt$\sim$}}
    \raise1pt\hbox{$<$}}}         
\def\gsim{\mathrel{\rlap{\lower4pt\hbox{\hskip1pt$\sim$}}
    \raise1pt\hbox{$>$}}}         

\title{Properties of the general NHDM. I. The orbit space}

\author{I.~P.~Ivanov$^{1,2}$, C.~Nishi$^{3}$
\\
  {\small $^1$ IFPA, Universit\'{e} de Li\`{e}ge, All\'{e}e du 6 Ao\^{u}t 17, b\^{a}timent B5a, 4000 Li\`{e}ge, Belgium}\\
  {\small $^2$ Sobolev Institute of Mathematics, Koptyug avenue 4, 630090, Novosibirsk, Russia}\\
  {\small $^3$ Universidade Federal do ABC,
	Rua Santa Ad\'elia, 166, 09.210-170, Santo Andr\'e, SP, Brazil}
}

\begin{document}
\maketitle

\begin{abstract}
We study the scalar sector of the general $N$-Higgs-doublet model via geometric constructions 
in the space of gauge orbits. We give a detailed description of the shape of the orbit space 
both for general $N$ and, in more detail, for $N=3$.
We also comment on remarkable analogies between NHDM and quantum information theory. 
\end{abstract}

\section{Introduction}

The Standard Model relies on the Higgs mechanism to realize the electroweak symmetry
breaking (EWSB).
Many variants of the Higgs mechanism have already been suggested,
but it is not presently known what particular variant is realized in Nature.

While waiting for the LHC to give the first hints of the Nature's choice, 
theorists must be duly prepared to safely interpret the future LHC data.
This implies, in particular, that theorists must be aware of all essential possibilities
which can be realized within a chosen model for EWSB. Since non-minimal Higgs sector
usually involve many free parameters, it is highly desirable to analyze the chosen model
in its most generic formulation, allowing for all possible degrees of freedom.
This general analysis of a specific model should show which phenomenological
consequences are universal and which are sensitive to values of the parameters,
which symmetries can in principle arise in the model and how they are broken,
which properties hold only at the tree-level and which survive the perturbation series.
When this general structure of a model is well understood, one should proceed further
and restrict the model by taking into account existing experimental constraints.

Unfortunately, such an exhaustive analysis is hardly feasible for many non-minimal Higgs sectors.
A very representative case is given by the 2HDM, \cite{TDLee,Hunter,CPNSh}.
Here, the straightforward algebra fails even at the very first step, because
the Higgs potential cannot be minimized explicitly in the general case.
As a result, for a long time only relatively simple variants of the 2HDM were analyzed,
while the most general 2HDM remained barely studied. 
In the last several years a number of tools were developed
which led to many insights into the properties of the general 2HDM.
These methods were based on the idea of the reparametrization symmetry, 
or basis-invariance, of the model: a unitary transformation between the Higgs doublets changes the parameters 
of the model, but nevertheless leads to the same physical properties of the observable particles.
This idea can be implemented via the tensorial formalism at the level of Higgs fields \cite{CP,haber,haber2,oneil} 
or via geometric constructions in the space of gauge-invariant bilinears 
\cite{sartori,nagel,heidelberg,nishi2006,ivanov0}.
In the latter case the formalism was extended to include non-unitary reparametrization 
transformations \cite{ivanov1,ivanov2,nishi2008},
which revealed interesting geometric properties of the 2HDM in the orbit space equipped
with the Minkowskian metric.

It is a natural idea to extend these successful techniques to $N$ doublets.
The general NHDM is obviously more involved than 2HDM, both at the level of
scalar sector and Yukawa interactions (see examples in
\cite{weinberg1976,Branco1980,DeshpandeHe1994}).
Some properties of the general NHDM were analyzed in 
\cite{Erdem1995,barroso2006,heidelberg,nishi:nhdm,ferreira-nhdm,zarrinkamar},
with a special emphasis on $CP$-violation, \cite{lavoura1994,nishi2006}.
However, a method to systematically explore all the possibilities offered with $N$ doublets
was still missing.

In principle, generalization from 2HDM to NHDM is straightforward in the tensorial formalism, 
however it is very difficult to translate tensorial invariants into physical observables.
On the other hand, the geometric approach in the space of bilinears offers a more appealing treatment
of the Higgs potential, but the shape of the NHDM orbit space is rather complicated
and has not been fully characterized so far.
In this paper we fill this gap by studying in detail the algebraic and geometric properties of
the NHDM orbit space.
Many of these results will be used in the companion paper \cite{ivanov2010} where the
minimization problem and the symmetry breaking patterns of the
Higgs potential of the general NHDM are analyzed.

The paper is organized as follows. Section 2 is devoted to three distinct, but interrelated 
approaches to description of gauge orbits in the space of Higgs fields.
Then, in Section 3 we construct the orbit space of NHDM as a certain algebraic manifold 
and discuss at length its algebraic and geometric properties. 
In Section 4 we treat the specific case of 3HDM in even greater detail, aiming not only 
at a concise algebraic description of the orbit space, but also trying to gain an intuitive understanding
of its shape. In the final Section we draw our conclusions.

\section{Describing Higgs fields in NHDM}\label{section:describing}

The scalar potential of the NHDM is constructed from gauge-invariant bilinear combinations
of the Higgs fields. The space of these combinations
can be described in three algebraically different but closely related ways:
via representative Higgs doublets, via a $K$-matrix, and via a vector in the adjoint space.
In this Section we describe and compare these three ways.  

\subsection{Field space and gauge orbits}\label{section:fields}

The scalar content of the general NHDM consists of $N$ complex Higgs doublets with electroweak isospin $Y=1/2$:
\be
\phi_a = \doublet{\phi_a^+}{\phi_a^0}\,,\quad a=1,\dots , N\,.
\ee
The total dimensionality of the space of scalar fields is $4N$.
Since the Higgs lagrangian is EW-symmetric, we can perform any simultaneous {\em intradoublet} $SU(2)\times U(1)$ transformation 
inside all doublets without changing the lagrangian.
If we take a generic point in the Higgs space and apply all possible EW transformations, we will get a four-dimensional manifold
called the (gauge) {\em orbit}. 
Thus, the entire $4N$-dimensional space of Higgs fields is naturally ``sliced'' into non-intersecting orbits.
The resulting set of orbits is a $(4N-4)$-dimensional manifold called the {\em orbit space}.

In principle, $\phi_a$ are operators. However, when minimizing the Higgs potential,
we will look for vacuum expectation values of the Higgs fields $\lr{\phi_a}$,
which are $c$-numbers. 
Then, we can characterize each orbit by a specific representative point in the Higgs space: 
\be
\label{Phi:generic}
\phi_1 = \doublet{0}{v_1}\,,\quad 
\phi_2 = \doublet{u_2}{v_2e^{i\xi_2}}\,,\quad 
\phi_a = \doublet{u_a e^{i\eta_a}}{v_a e^{i\xi_a}}\,,\quad a > 2\,.
\ee
This point (and therefore, the entire orbit) is characterized by $4N-4$ real parameters: $N$ values of $v_a$,
$N-1$ values of $u_a$, $a>1$, $N-1$ phases $\xi_a$, $a>1$, and $N-2$ phases $\eta_a$, $a>2$. 

It is well known that if at least one $u_a \not = 0$, such a point corresponds to the {\em charge-breaking vacuum},
in which the electroweak symmetry is broken completely and the photon acquires mass.
If we insist that the vacuum be neutral, we must set all $u_a = 0$.
Thus, the representative point of a generic {\em neutral orbit} is
\be
\label{Phi:neutral}
\phi_1 = \doublet{0}{v_1}\,,\quad 
\phi_a = \doublet{0}{v_a e^{i\xi_a}}\,,\quad a > 1\,.
\ee
It is characterized by $N$ parameters $v_a$ and $N-1$ phases $\xi_a$, making the dimensionality of the {\em neutral orbit space}
equal to $2N-1$, which is $2N-3$ units less than the dimensionality of the entire orbit space.

\subsection{Reparametrization freedom}

So far we have mentioned only the electroweak $SU(2)\times U(1)$ transformations between the components
of each doublet. Since the potential is an EW scalar, such transformation do not affect the parameters of the potential.

Now consider the $SU(N)_H$ group of transformations that mix the doublets without affecting their intradoublet structure
(index $H$ stands for ``horizontal''). This transformation sends a given Higgs potential
to another viable Higgs potential with different coefficients. Such a transformation is called
a {\em reparametrization transformation}, or a {\em horizontal space transformation}, or a {\em Higgs-basis change}.
The key property of this transformation is that although it reparametrizes the potential,
it leaves the physical observables invariant \cite{CP,haber}. 
This property is known as the reparametrization invariance,
or the Higgs-basis invariance, of the model.
The same is true for anti-unitary transformations as well, so one can state that the reparametrization
group of the general NHDM consists of all unitary and anti-unitary transformations acting in the space $\mathbb{C}^N$:
$\phi_a \to U_{ab}\phi_b$ and $\phi_a \to U_{ab}\phi^*_b$. The antiunitary
transformations are also known as generalized
CP transformations\,\cite{ecker-grimus,HaberSilva}.

Reparametrization transformations link different gauge orbits. If we pick up a specific point in the gauge orbit space, 
then by applying all possible reparametrization transformations
we can reach many other points in the orbit space. Thus, the orbit space itself becomes split into non-intersecting $SU(N)$-orbits.
We will study this stratification in more detail in Section~\ref{section:isotropy}. 

For any neutral orbit parametrized by $v_a$, $\xi_a$ according to (\ref{Phi:neutral}),
we can find a reparametrization transformation that brings it to a ``canonical
form'' (also known as \textit{Higgs basis})
\be
\label{Phi:neutral:canonical}
\phi_1 = \doublet{0}{v}\,,\quad 
\phi_a = \doublet{0}{0}\ \mbox{for}\ a > 1\,, \quad v^2 \equiv \sum_a v_a^2\,.
\ee
Obviously, we have $N$ equivalent canonical forms depending on which doublet has the
non-zero value. These equivalent forms can be related to each other by a discrete
subgroup (permutation) of the reparametrization group.

Equivalently, any point in the charge-breaking orbit space can be brought to its own
canonical form
\be
\label{Phi:generic:canonical}
\phi_1 = \doublet{0}{v}\,,\quad 
\phi_2 = \doublet{u}{0}\,,\quad 
\phi_a = \doublet{0}{0}\ \mbox{for}\ a > 2\,,\quad
v^2 \equiv \sum_a v_a^2\,, \ u^2 \equiv \sum_a u_a^2\,.
\ee
There are $N(N-1)/2$ such canonical choices.
To avoid double counting of equivalent canonical forms within such choices we can
restrict $v^2\ge u^2$.
The neutral orbit space corresponds to the limit $u^2\to 0$.
The other extremum, $u^2=v^2$, corresponds to a rather special space
of ``{\em maximally charge-breaking}'' (MCB) vacua.

\subsection{$K$-matrix formalism}

Following \cite{nagel,nishi2006,heidelberg}, we represent the electroweak-scalar bilinears defined above 
as components of a complex hermitean $N\times N$ matrix:
\be
\label{K:def}
K_{ab} \equiv (\phi^\dagger_b\phi_a)\,.
\ee
Several important properties of the $K$-matrix were proven in
\cite{heidelberg,nishi:nhdm}:
\begin{itemize}
\item
It is hermitean positive-semidefinite matrix.
\item
Its rank is two for a generic (charge-breaking) vacuum and one for a neutral vacuum. 
This result stems from the fact that we deal with electroweak {\em doublets} and not higher representations
of the gauge group.
\end{itemize}
In other words, the $K$-matrix has at most two non-zero (and positive) eigenvalues, while the other $N-2$ eigenvalues are zero.
For the neutral vacuum, one gets only one non-zero (positive) eigenvalue and $N-1$
zeros.
The maximally charge-breaking vacua can be defined in terms of
$K$-matrices by $[\Tr K]^2 = 2 \Tr(K^2)$.

A reparametrization transformation $U\in SU(N)_H$ acting on doublets $\phi$ transform
also the $K$-matrix according to the adjoint transformation law:
\eq{
\label{K:U}
K \to UKU^{\dag}.
}
In particular, starting from any $K$-matrix, one can always find such a
transformation that
diagonalizes it:
\bea
\label{K:canonical}
K &=& \mathrm{diag}(v^2,0,\dots,0)\quad \mbox{for neutral orbit space;}\nonumber\\
K &=& \mathrm{diag}(v^2,u^2,\dots,0)\quad \mbox{for charge-breaking orbit space.}
\eea
These diagonal $K$-matrices correspond to the ``canonical'' orbits (\ref{Phi:neutral:canonical})
and (\ref{Phi:generic:canonical}). 

From the observation that the $K$-matrix is a hermitean and rank 2 matrix, we
can also deduce the dimensions of the neutral and charge breaking orbit space. Let us
suppose rank$(K)=2$. Then among the $N$ lines (columns) of $K$, there are at most two 
linearly independent. We can arbitrarily choose them to be the first
and second lines (columns) by relabeling the doublets. The remaining
lines (columns) can be rewritten as linear combinations of lines 1 and 2, and,
because of hermiticity, the expansion coefficients are not arbitrary but determined
by
the elements of the linearly independent lines. This fact is proved in
appendix\,\ref{ap:maximal}. Therefore, we can choose the set
\eqarr{
\label{minimal:K}
K_{1a}&=&\phi^\dag_a\phi_1, \quad a=1,\ldots,N\,,\cr
K_{2b}&=&\phi^\dag_b\phi_2, \quad b=2,\ldots,N\,,
}
as the set of $4N-4$ algebraically independent gauge invariants
$K_{ab}=\phi^\dag_b\phi_a$, noticing that $K_{11}$ and $K_{22}$ are real while the
rest are complex. Thus the dimension of the charge breaking orbit space is $4N-4$.
The neutral orbit space corresponds to the subcase of rank$(K)=1$ where we can choose
the first line (column) to be the linearly independent line (column), which
implies that the dimension of the neutral orbit space is $2N-1$.
The dimensions of the neutral and charge-breaking orbit spaces can be also
deduced using the isotropy groups acting on $K$; see details in
Section~\ref{section:isotropy} below.

\subsection{Adjoint representation}\label{section:adjoint}

Yet another look at the orbit space of the $N$-Higgs-doublet model is offered by 
the adjoint representation of the reparametrization group $SU(N)_H$.

Since the $K$-matrix (\ref{K:def}) is a hermitean $N\times N$ matrix, it can be
decomposed as
\be
\label{K:decomposed}
K \equiv r_0 \cdot \ml{\sqrt{\!{2 \over N(N-1)}}}\,{\bf 1}_N + r_i \lambda_i\,,
\quad i = 1, \dots , N^2-1\,.
\ee
Here $\lambda_i$ are generators of $SU(N)$ satisfying relations
$$
\lambda_i\lambda_j = {2\over N}\delta_{ij} {\bf 1}_N + i f_{ijk}\lambda_k + d_{ijk}\lambda_k\,.
$$
The coefficient in front of the unit matrix in (\ref{K:decomposed}) is chosen for
future convenience. The values of $r_0$ and $r_i$ can be extracted from the
$K$-matrix:
\be
r_0 = \ml{\sqrt{\!{N-1\over 2N}}}\Tr\,K\,,\quad r_i = {1\over 2}\Tr[K\lambda_i]\,.
\ee

If the space of $K$-matrices were generalized to the space of all $N\times N$
hermitean matrices, i.e., $M_h(N)$, Eq.~(\ref{K:decomposed}) would define a linear
and invertible map between $M_h(N)$ and the space of all possible real vectors $r^\mu
\equiv (r_0, r_i)$, i.e., $\RR^{N^2}$\,\cite{nishi:nhdm}, which we call the {\em
adjoint space}.
The notation $r^\mu$ alludes to the Minkowski-space formalism similar to what was 
developed for 2HDM in \cite{ivanov1,ivanov2,nishi2006,nishi2008}.
Since we limit ourselves only to the (anti)unitary reparametrization transformations,
we are not going to use this formalism here, although we do think that it might be useful in NHDM.
Here we use $r^\mu$ just as a short notation of $(r_0,r_i)$.

When $M_h(N)$ is restricted to the space of positive semidefinite rank 2
$K$-matrices, which we call the \textit{$K$-space}, the mapping \eqref{K:decomposed}
from the $K$-space to the adjoint space $\mathbb{R}^{N^2}$ is no longer surjective
and its image defines a manifold embedded in $\mathbb{R}^{N^2}$
which will be denoted by $\calV_\Phi$\,\cite{nishi:nhdm} (orbit space). 

The shape of $\calV_\Phi$, which we study in Section~\ref{section:NHDM-orbit-space} below,
is rather complicated. However, to cast the first glance at it, let us remind the reader 
of the situation in 2HDM. There, the $K$-matrix 
was a hermitean 2-by-2 matrix, and the condition rank$K \le 2$ was automatically
satisfied.
The positive-semidefiniteness led to 
\be
\label{K:conditions:2HDM}
\Tr K \ge 0\,,\quad (\Tr K)^2 - \Tr[K^2] \ge 0\,,
\ee
which in the adjoint representation translated into
\be
\label{LC:def}
r_0\ge 0\,,\quad r_0^2 \ge \vec r^2\,,
\ee
where $\vec r$ is just an index-free label for $r_i$.
Thus, in 2HDM the orbit space was represented by (the surface and the interior of)
the forward lightcone in $\mathbb{R}^4$.

With more than two doublets, the rank-2 requirement generates extra conditions 
to be imposed in addition to (\ref{K:conditions:2HDM}).
These conditions are formulated as equalities (not inequalities) among traces of powers of the $K$-matrix up to $K^N$,
see details in Section~\ref{section:NHDM-orbit-space}.
In the adjoint space they will translate into a set of algebraic equations on components of $r^\mu$. 
The orbit space still lies on and inside the forward lightcone (\ref{LC:def}), 
but it occupies only a certain region inside it. 

A unitary reparametrization transformation of $\phi_a$ keeps $r_0$ unchanged 
but leads to a rotation of the vector $\vec r$ in $\mathbb{R}^{N^2-1}$.
An anti-unitary reparametrization transformation adds to that a reflection 
of $N(N-1)/2$ components of $\vec r$.
Since the map $SU(N) \to SO(N^2-1)$ is not surjective, not all rotations of $\vec r$ 
can be induced by a reparametrization transformation of the doublets. 
Thus, the reparametrization group in the adjoint space is a certain (and rather small) subgroup
of the full rotation group $O(N^2-1)$.
This is also reflected in the rather complicated shape of the orbit space itself, 
which is invariant only under rather special rotations.

As discussed above, the $K$-matrix of any canonical neutral (\ref{Phi:neutral:canonical})
or charge-breaking (\ref{Phi:generic:canonical}) orbit is diagonal. Its decomposition (\ref{K:decomposed})
involves only diagonal matrices $\lambda_i$, i.e. the Cartan subalgebra of $su(N)$ together with the unit matrix.
Thus, the canonical orbits correspond in adjoint space to a certain $(N-1)$-dimensional section through the {\em root space} 
of $su(N)$, to be discussed in Section~\ref{subsection:root-space}. 

\subsection{Higgs potential}

Let us also discuss how the Higgs potential is described within each of these approaches.

The generic Higgs potential in NHDM can be written in a tensorial form \cite{CP,haber}:
\be
\label{V:tensorial}
V = Y_{\bar{a}b}(\phi^\dagger_a \phi_b) + Z_{\bar{a}b\bar{c}d}(\phi^\dagger_a \phi_b)(\phi^\dagger_c \phi_d)\,,
\ee
where all indices run from 1 to $N$.
The potential is constructed from $N^2$ bilinears $(\phi^\dagger_a \phi_b)$, and therefore it can be viewed as defined in
the orbit space\footnote{The idea to switch to the orbit space in order to simplify the task of a group-invariant potential minimization
is rather old, see \cite{sartori-old} and references therein.}.
Coefficients in the quadratic and quartic parts of the potential are grouped into components of tensors $Y_{\bar{a}b}$
and $Z_{\bar{a}b\bar{c}d}$, respectively;
there are $N^2$ independent components in $Y$ and $N^2(N^2+1)/2$ independent components in $Z$.

Within the $K$-matrix approach, the potential is still based on the same tensors and can be written symbolically as
\be
\label{V:Kmatrix}
V = \Tr[YK]+\Tr\Tr[ZK\otimes K]\,,
\ee
where ``$\Tr\Tr$'' indicates the traces over the two pairs of indices.

In the adjoint space, one replaces a pair of doublet indices by the index $\mu = (0,i)$.
The Higgs potential take the following form: 
\be
\label{V:general}
V = -M_\mu r^\mu + {1\over 2}\Lambda_{\mu\nu}r^\mu r^\nu \equiv 
- (M_0 r_0 + M_i r_i) + {1\over 2}\left(\Lambda_{00}r_0^2 + 2 \Lambda_{0i}r_0 r_i + \Lambda_{ij}r_ir_j\right)\,.
\ee
The scalar $M_0$ and the vector $M_i$ are essentially $Y_{\bar{a}b}$ of (\ref{V:tensorial}) written in the adjoint space.
The total number of free parameters in $M_0$ and the vector $M_i$ is $1+ (N^2-1) = N^2$.
The scalar $\Lambda_{00}$, vector $\Lambda_{0i}$ and symmetric tensor $\Lambda_{ij}$ represent $Z_{\bar{a}b\bar{c}d}$
in the adjoint space; their parameter counting gives $1+(N^2-1)+(N^2-1)N^2/2 = (N^2+1)N^2/2$.
We stress again that we do not use Minkowskian metric in this paper, so all the contractions 
of pairs of indices are understood in the Euclidean sense. 

Let us discuss the relation among the three ways of describing the Higgs field configurations in NHDM.

Working in terms of fields parametrized as (\ref{Phi:generic}) and (\ref{Phi:neutral}),
one can easily describe the entire orbit space available in terms of $v_a$, $u_a$ and relative phases.
The price one pays for this simplicity is that the Higgs potential involves tensors
$Y_{\bar{a}b}$ and $Z_{\bar{a}b\bar{c}d}$, whose properties are very far from being intuitive.
For example, even in 2HDM one must resort to long computer-assisted algebraic
manipulation in order to formulate the explicit $CP$-invariance of the
potential\,\cite{haber2}.
One can expect that understanding the Higgs potential of NHDM will be even harder.

In the adjoint $r^\mu$ space the description of the orbit space becomes much more complicated. 
On the other hand, the treatment of the Higgs potential is dramatically simpler.
For example, in 2HDM one could easily formulate and prove conditions for existence of a symmetry, 
one could derive theorems about the number and coexistence of minima, etc. 
Remarkably, the same treatment can be extended straightforwardly to NHDM, which will be the subject 
of the companion paper \cite{ivanov2010}. 
Thus, the orbit space description represents the only essential complication on the way to understanding 
the properties of the scalar sector in generic NHDM.

The $K$-matrix formalism lies somewhere in between. The $K$-space resembles $\calV_\Phi$ in the adjoint space,
but the Higgs potential is still written in a non-intuitive tensorial form.
However, many lines of argumentation can be started at the level of $K$-matrix.

\subsection{NHDM as a quantum information-theoretic problem}

It is a remarkable and perhaps an underappreciated fact that the mathematics 
behind constructing the orbit space in the $N$-Higgs-doublet model
is very similar to what is studied in the Quantum Information Theory
(for introduction, see recent textbooks \cite{book-qnetworks,book-geometry}) 
and in the so-called geometric quantum mechanics, \cite{geometric-QM}. 

In quantum information theory one studies quantum evolution of interacting $N$-level quantum states, called {\em qudits}
(for $N=2$ and $N=3$ one speaks of qubits and qutrits, respectively), which are not necessarily isolated
from the environment. 
One of the basic problems here is to describe the space of states of a single $N$-level qudit.
In general, it is described by a hermitean positive-semidefinite $N\times N$ density
matrix $\hat\rho$,
which satisfies certain axioms, \cite{Boya-grouptheory,book-qnetworks,book-geometry}.
For pure states $\mathrm{rank}\hat\rho = 1$, while for a mixed state $ 1 < \mathrm{rank}\hat\rho \le N$.
The density matrix can also be decomposed in the basis of the algebra $su(N)$, similarly to (\ref{K:decomposed}):
\be
\label{rho:decomposed}
\hat \rho = {1 \over N} {\bf 1}_N + \rho_i \lambda_i\,, \quad i = 1, \dots , N^2-1\,,
\ee
where the coefficient in front of the unit matrix is fixed by condition $\Tr\hat\rho = 1$.
The vector $\vec \rho$ is known as the {\em coherence vector}, or the {\em Bloch vector}.
All possible Bloch vectors occupy a region in the $(N^2-1)$-dimensional space called 
the (generalized) Bloch ball. 
It is remarkable that only recently the structure of the Bloch ball was analyzed for $N>2$,
\cite{Kimura,Byrd,Kossakowski2003,book-geometry}.

Many of these objects have counterparts in HNDM.
The density matrix corresponds to an appropriately normalized $K$-matrix. 
The neutral vacuum of NHDM corresponds to a pure states of a qudit,
while the charge-breaking vacuum corresponds to a mixed qudit state
with a rank-2 density matrix. Higher rank density matrices do not have their counterparts in NHDM.
The coherence vector for the $N$-qudit is analogous to the adjoint space
vector $\vec r$, and the generalized Bloch sphere is then just another name for the gauge orbit space.

It is well possible that the analogy between the quantum information theory and NHDM
can be pursued further. In any case, we believe that the elaborate mathematics of quantum theory 
can generate new insights into the properties of NHDM or similar problems in particle physics.

\section{Orbit space of NHDM}\label{section:NHDM-orbit-space}

\subsection{Geometric description of the orbit space}
\label{subsec:geometric}

Let us start by describing some geometric properties of the orbit space of NHDM $\calV_\Phi$ in the adjoint 
space of vectors $r^\mu$, which can be inferred directly from the definitions. 

Let us first note that the orbit space has a conical shape: if point $r^\mu \in \calV_\Phi$, then
$\alpha r^\mu \in \calV_\Phi$ for any $\alpha \ge 0$. Therefore, in order to understand the shape of 
$\calV_\Phi$, it is sufficient to study its $r_0 = \mathrm{const}$ section at any positive $r_0$.
To this purpose, we switch to the $(N^2-1)$-dimensional space of normalized vectors $n_i \equiv r_i/r_0$.
The neutral orbit space then lies on the surface of the unit sphere $\vec n^2 = 1$, while
the charge-breaking orbit space occupies a region strictly inside it.

It is plain to see that a point in the neutral orbit space in the $\vec n$-space is parametrized
by $N$ independent complex numbers up to an overall (complex) factor. In other words,
the points of the neutral orbit space are in one-to-one correspondence with complex rays
in $\mathbb{C}^N$ passing through the origin, which form the $(N-1)$-complex-dimensional complex projective
space $\mathbb{CP}^{N-1}$. Thus, the neutral orbit space in the $\vec n$-space has 
the shape of $\mathbb{CP}^{N-1}$ embedded into $\mathbb{R}^{N^2-1}$.

The entire orbit space of NHDM can be reconstructed from the neutral orbit space by an operation, 
which we call {\em self-join}. By definition, the self-join of a set of points $S$ in an affine space
is a union of points lying on straight line segments drawn between all pairs of points of $S$.
Now, let us pick up two points from the neutral orbit space, whose $K$-matrices $K_1$ and $K_2$ 
are not proportional to each other.
Consider the open interval of $K$-matrices lying between them:
\be
K = c K_1 + (1-c) K_2\,,\quad c \in (0,1)\,.
\ee
Such $K$-matrices are necessarily rank-2 matrices. Inversely, any rank-2 $K$-matrix can be written 
(not uniquely) as a linear superposition with positive weights of a pair of rank-1 $K$-matrices.
Since the map from the $K$-space to the orbit space is linear, 
the same construction holds in the $r^\mu$-space, which proves that the entire orbit space is a self-join
of the neutral orbit space.
In loose terms, the charge-breaking orbit space is ``stretched'' on the wireframe of the neutral orbit space.

Note that this construction is similar, but not completely analogous, to the convex hull 
that arises in the quantum information theory, where the density matrices can have an arbitrary rank.
It also means that the NHDM orbit space does not possess the strict convexity.

\subsection{Algebraic description of the orbit space}

At the level of $K$-matrix, the defining criterion is that the $K$-matrix is a positive-semidefinite matrix with rank $\le 2$
\cite{heidelberg,nishi:nhdm}. In other words, it requires that there be at most
two non-zero eigenvalues, which must be positive.
Following \cite{Kimura,Byrd}, we write the characteristic equation for the $K$-matrix
as
\be
\label{characteristic:N}
\mathrm{det}(\lambda {\bf 1}-K) = \lambda^N +
\sum_{j=1}^{N}(-1)^{j}s_j\ms{(K)}\lambda^{N-j}\,.
\ee
The coefficients $s_k$ can be written as products of the eigenvalues $\lambda_i$,
i.e., the roots of the characteristic equation,
\be
s_n = \sum_{1 \le i_1 < \cdots < i_n \le N} \prod_{j=1}^n \lambda_{i_j}\,,
\ee
as well as in terms of traces of powers of the $K$-matrix:
\eq{
\label{sn:def}
s_n\ms{(K)} = \frac{\ms{(-1)^{n-1}}}{n}
\Tr\big[K^n+\sum_{j=1}^{n-1}(-1)^{j}s_{j}\ms{(K)}K^{n-j}\big]
\,,~~n=1,\ldots,N\,.
}
For example, $s_1\ms{(K)} = \Tr[K]$ and
$s_2\ms{(K)} = {1\over 2} \left(\Tr^2[K] -\Tr[K^2]\right)$.
When written without variables, the identification $s_n\equiv s_n(K)$ will be
assumed.
In general, positive semi-definiteness of matrix $K$ is equivalent to non-negative
values of all $s_n$.
In our case, the requirement that the $K$-matrix has rank $\le 2$ is equivalent to
\be
\label{Sa}
s_1\ms{(K)} \ge 0\,, \quad s_2\ms{(K)} \ge 0\,, \quad s_n\ms{(K)} = 0 \quad \mbox{for
all}\quad 2 < n \le N\,.
\ee
Since $K$ is hermitean and hence diagonalizable, the minimal annihilating polynomial
is, instead of Eq.\,\eqref{characteristic:N},
\eq{
K\big[K^2-s_1\ms{(K)}K+s_2\ms{(K)}\bs{1}\big]=0\,,
}
which automatically guarantees Eq.\,\eqref{Sa}.
For the neutral orbit space we require that there be only one positive eigenvalue,
i.\,e.,
\be
\label{Km:inequalities2}
s_1\ms{(K)} \ge 0\,, \quad s_n\ms{(K)} = 0\,, \quad
\mbox{for all}\quad 2 \le n \le N\,,
\ee
which can be summarized by
\eq{
K^2=s_1\ms{(K)}K\,.
}

From Eqs.\,\eqref{characteristic:N} or \eqref{sn:def}, it is clear that the
coefficients $s_n\ms{(K)}$ are functions of $K$ invariant by the reparametrization
group action in Eq.\,\eqref{K:U}. Because of the positive semi-definiteness of $K$
and rank$(K)\le2$, such action divides the space $\calV_\Phi\subset\RR^{N^2}$ into
$SU(N)$ orbits. Each of these orbits can be uniquely characterized by the set
$\{s_1,s_2\}$ of $SU(N)$ invariants (reparametrization invariants),
since the other $s_n$ are null. If $K$ were allowed to be a general hermitian
matrix, then all the set $\{s_n\}$ of $N$ invariants would be necessary to
characterize all the orbits. The one-to-one correspondence between an orbit and a
set of invariants applies because they uniquely define the eigenvalues of the matrix
$K$, in a given order, and the set of all matrices with the same eigenvalues are
conjugated to the same
diagonalized matrix, forming then one orbit.
The invariants can be calculated and written in terms of $r^\mu$ in
Eq.\,\eqref{K:decomposed}:
\eqarr{
\label{s:1}
s_1\ms{(K)}&=&
\ml{\sqrt{\!\frac{2N}{N-1}}}\,r_0\,,
\\
\label{s:2}
s_2\ms{(K)}&=&
r^2_0-\vec{r} ^2\,,
\\
\label{s:3}
s_3\ms{(K)}&=&
\ml{\frac{2}{3}}d_{ijk}r_ir_jr_k
-\ml{\frac{2(N-2)}{\sqrt{2N(N-1)}}}\,r_0[\vr^2-\frac{r_0^2}{3}]\,,
\\
\label{s:4}
s_4\ms{(K)}&=&
-\ml{\frac{1}{2}}\Gamma^{(4)}_{ijkl}r_ir_jr_kr_l
+\ml{\frac{1}{2}}\vr^4
+\ml{\frac{(N-2)(N-3)}{N(N-1)}}\,r^2_0[\vr^2-\frac{r_0^2}{2}]\,,
\\
\vdots ~\quad&& \hspace{3em}\vdots
\nonumber
}
where we defined the totally symmetric tensors
\eq{
\label{Gamma:n}
\Gamma^{(n)}_{i_1i_2\cdots i_n}\equiv
\frac{1}{2(n!)}\Tr[\{\lambda_{i_1}\cdots\lambda_{i_n}\}_+]\,.
}
The symbol $\{~\}_+$ denotes the sum of strings of $\lambda$'s with all possible
permutations of	 indices $i_1$ to $i_n$.
We can easily identify $\Gamma^{(2)}_{ij}=\delta_{ij}$ and
$\Gamma^{(3)}_{ijk}=d_{ijk}$.
Notice Eq.\,\eqref{s:4} already assumes $s_3=0$ in Eq.\,\eqref{s:3}.
Therefore, we can define the orbit space $\calV_\Phi$ as the set of
vectors $r^\mu\in\RR^{N^2}$ that satisfy the set of equalities and inequalities of
Eq.\eqref{Sa}, explicitly given by
\be
\label{sun:constraints}
r_0 \ge 0\,, \quad r_0^2 - \vec r^2 \ge 0\,,\quad
d_{ijk}r_ir_jr_k + \ml{N-2 \over\sqrt{2N(N-1)}}\,r_0(r_0^2-3\vec r^2) = 0\,,
\quad\cdots ~~,
\ee
where each equality $s_n=0$, $n\ge 3$, gives an algebraic equation of order $n$ in
$r^\mu$.
A systematic procedure to find the the higher order equations $s_n=0$ is given in
appendix \ref{ap:invariants}.
In general, $s_n\ms{(K)}=0$ is equivalent to
\eq{
\label{sn=0}
s_n\ms{(r_i\lambda_i)}-
\ms{(-1)^{n}}
\ml{\binom{\!N-2}{n-2}}
\Big(\!\frac{s_1}{\ms{N}}\Big)^{n-2}
[\vr^2-r^2_0(1-\ml{\frac{2}{n}})]=0
\,,\quad n\ge 2\,,
}
where the first term depends only on $\vr$ and it can be written as a sum of terms
containing contractions of the tensors in Eq.\,\eqref{Gamma:n} up to order $n$, see
appendix \ref{ap:invariants}.
We can see that, for each set of values of the invariants $r_0$ and $\vec{r}^2$,
Eqs.\,\eqref{sun:constraints} and \eqref{sn=0} lead to a set of $N$ algebraic
equations of order $n\le N$ that defines a manifold on $\RR^{N^2}$. Each of these
manifolds constitutes a single $SU(N)$ orbit in the orbit space because all the
available invariants $s_1,s_2$, are fixed.
In particular, the neutral orbit space is a particular $SU(N)$ orbit for which 
the second condition of Eq.\eqref{sun:constraints} becomes an equality:
\be
\label{LCcondition}
r_0^2 - \vec r^2 = 0\,,
\ee
meaning that the neutral orbit space must lie on the forward lightcone.

For a complete characterization of the orbits, it remains to specify the
range of variation for $\{r_0,\vec{r}^2\}$.
There is no upper bound for $r_0$. Let us take a fixed non-null value for $r_0$.
Then, $\vec{r}^2$ can decrease continuously from $\vec{r}^2=r^2_0$ to a lower bound
given by
\eq{
(\vec{r}^2)_{\rm min}=r^2_0-(s_2)_{\rm max}\,,
}
where $(s_2)_{\rm max}$ can be calculated using a diagonal matrix
$K=\diag(x_+,x_-,0,\cdots,0)$. We have to maximize $s_2=x_+x_-$ subjected to
$s_1=x_+ +x_-=\text{const}$ and $x_+,x_-\ge0$. We easily find that $x_+=x_-=s_1/2$
maximizes $s_2$ which yields, using Eq.\,\eqref{s:1},
\eq{
\label{nmin}
\frac{(\vec{r}^2)_{\rm min}}{r^2_0}=
a^2_N \equiv
\frac{N-2}{2(N-1)}\,.
}
In fact, we can write the two non-null eigenvalues $x_{\pm}$ of $K$ as
\eq{
x_{\pm}=\ml{\sqrt{\!\frac{N}{2(N-1)}}}\,r_0
\pm \sqrt{\vec{r}^2-a^2_N r^2_0}
\,.
}
Therefore, the $SU(N)$ orbits are entirely specified by the value of
$\vec{n}^2=\vec{r}^2/r^2_0\in [a_N,1]$. 
In particular, $\vr^2=a^2_Nr^2_0$, instead of Eq.\,\eqref{LCcondition}, defines the
maximally charge breaking space.

The result (\ref{nmin}) implies that, in the $r^\mu$-space, 
the orbit space is restricted to a conical region between two coaxial cones: 
the lightcone, and the inner one defined by (\ref{nmin}).
This observation might lead to non-trivial topological properties of the model,
similar to what was described in \cite{ivanov1}.

Finally, when passing from the full to the neutral orbit space, 
one might be surprised that a single algebraic condition (\ref{LCcondition})
reduces the dimensionality by $2N-3$ units, from $4(N-1)$ (charge-breaking) 
to $2N-1$ (neutral).
To show how this happens, let us write $s_2$ in terms of doublets. We find
\be
\label{s2:zab}
s_2\ms{(K)} = \sum_{1\le a<b\le N} z_{ab}\,,\quad 
\mbox{where}\quad
z_{ab} \equiv (\phi_a^\dagger \phi_a)(\phi_b^\dagger \phi_b) - (\phi_a^\dagger
\phi_b)(\phi_b^\dagger \phi_a)\,.
\ee
There are $N(N-1)/2$ quantities $z_{ab}$, and each of them is non-negative due to the Schwarz lemma.
Not all of $z_{ab}$ are independent, though. Suppose that all norms $(\phi_a^\dagger \phi_a)$ are fixed.
Then, for the first three doublets, the quantities $z_{12}$, $z_{13}$ and $z_{23}$ are algebraically
independent. However for any extra doublet, e.g. $\phi_q$, only two of $z$'s, 
e.g. $z_{1q}, z_{2q}$, can be chosen independently.
Any further $z_{aq}$ with $a>2$ is not independent anymore but is linked to previous
$z$'s by an algebraic relation,
see a proof in Appendix~\ref{appendix:zab}.
This is a consequence of the fact that we deal with doublets, not higher representations
of the gauge group. Thus, for $N$ doublets we have $2N-3$ independent $z_{ab}$.
Now, requiring that $s_2=0$ automatically sets {\em all} $z_{ab}=0$, which means that
it is equivalent to $2N-3$ independent equalities.

\subsection{Root space}\label{subsection:root-space}

Reparametrization transformation of the doublets, $\phi_a \to \bar\phi_a = U_{ab}\phi_b$, described by a unitary matrix $U_{ab} \in SU(N)$
corresponds in the adjoint space to a certain rotation of the vector $\vec r$: $r_i
\to \bar r_i = O_{ij}r_j$, where $O_{ij}=O_{ij}(U)$.
The transformation matrix $O_{ij}$ belongs to the group adj$SU(N)$
(adjoint representation), which is only a proper subgroup of $SO(N^2-1)$. It means that
not all rotations in $SO(N^2-1)$ can be induced by reparametrization
transformations.

This fact restricts the way we can manipulate the adjoint orbit space. However, we
always have a reparametrization freedom to bring any initial $K$-matrix to the
diagonal form. In the adjoint space, it corresponds to certain allowed rotations of
the entire orbit space that bring any point down to the $(N-1)$-dimensional
\textit{root space}, which describes the diagonal $K$-matrices. In the $\vec n$
space the $N$ neutral orbits are represented by vertices of a regular
$(N-1)$-simplex, while the charge-breaking orbit space is represented by the edges
of this simplex, i.e. by the line segments joining the vertices. This gives the full
description of the orbit space in the root space.
For example, the orbit space restricted to the root space corresponds, for $N=3$, to
the vertices and edges of an equilateral triangle while, for $N=4$, it corresponds to
the vertices and edges of a regular tetrahedron.
The case $N=3$ will be treated in more detail in Sec.\,\ref{sec:3hdm}.

The vectors in the root space can be parametrized in a very symmetric way in terms of $N$ 
barycentric coordinates $p_i$ constrained by $\sum_{i=1}^{N}p_i=1$ and
$p_i\ge 0$:
\eq{
\frac{K}{r_0}=
\ml{\sqrt{\!\frac{2N}{N-1}}}\,\diag(p_1,p_2,\cdots,p_N) = 
\ml{\sqrt{\!\frac{2}{N(N-1)}}}\,\id+
\sum_{i=1}^{N}p_iq_i\,,
}
where we defined the traceless matrices
\eq{
q_i\equiv
\ml{\sqrt{\!\frac{2N}{N-1}}}\big(e_{ii}-\frac{\id}{N}\big)\,
}
where $e_{ii}$ are the canonical matrices defined by
$(e_{ij})_{kl}=\delta_{ik}\delta_{jl}$.
Additionally, we can have at most two non-null $p_i$, since rank $K\le 2$.
The vertices of the simplex, corresponding to the neutral orbit, are given by
$\vec{p}=(1,0,\ldots,0), (0,1,\ldots,0),\ldots,(0,\ldots,0,1)$. Notice the matrices
$q_i$ are not all independent but obey $\sum_{i=1}^N q_i=\bs{0}$. Various geometric
features can be calculated explicitly by using the coordinates $p_i$.

The orbit space in the root space has a residual discrete symmetry with group 
$S_N$, related to the permutations of the doublets and corresponding permutations of
the vertices of the simplex. Thanks to this freedom, we conclude that {\em any}
neutral orbit can be brought to a predefined vertex, which means that all the points
in the neutral orbit space are conjugate to each other, that is, can be mapped to
each other by a reparametrization transformation. As for the charge-breaking points,
one can always use the reparametrization freedom to place it on any predefined edge
of the simplex, and even more, on any of the two symmetric halves of the edge.
Therefore, if we are restricted to the points not conjugated by $S_N$, we get a line
segment going from one vertex to the middle point of an edge. Such minimal set is in
one-to-one correspondence to the $SU(N)$ orbits in the $\vec{n}$-space.
As we already discussed, one parameter can be chosen to characterize each
point in the line segment, i.e., $|\vec{n}|\in [a_N,1]$.
Then the whole orbit space $\calV_\Phi$ can be recovered by $SU(N)$ conjugation on
this line segment and by varying $r_0$.
One can also recover the result (\ref{nmin}) just from the shape 
of the orbit space in the root space by calculating the 
distance from the midpoint of an edge of the $(N-1)$-simplex to its center.

\subsection{Isotropy groups and $SU(N)$ stratification}\label{section:isotropy}

Let us also describe the isotropy groups of the charge-breaking and neutral vacua, that is,
the subgroups of the total reparametrization group that leave invariant a given point.

Let us take a point in the neutral orbit space and bring it down to the root space,
turning its $K$-matrix into diag$(v^2,0,\dots,0)$.
It remains invariant under any $U(N-1)$ transformation that does not involve the first doublet
as well as under a $U(1)$ phase rotation of the first doublet alone. 
Since the bilinear are insensitive to the overall phase rotation,
we get the isotropy group of the neutral vacuum $SU(N-1)\times U(1)$, 
which is a $(N-1)^2$-dimensional Lie group.

Since the entire reparametrization group $SU(N)$ has $N^2-1$ parameters, 
there are $2N-2$ generators that do shift
a chosen point along mutually orthogonal directions in the orbit space.
Therefore, the neutral orbit space has $2N-2$ dimensions in the $\vec n$ space, 
and $2N-1$ dimensions in the $r^\mu$ space.
This coincides with the calculations of Section~\ref{section:fields}.

Now take a generic point in the charge-breaking orbit space, with its $K$-matrix
being \linebreak diag$(v_1^2,v^2_2,0,\dots,0)$. It remains invariant under $U(N-2)$
transformation of the last $N-2$ doublets
as well as phase rotations of the first two doublets.
The isotropy group is therefore $SU(N-2)\times U(1) \times U(1)$, whose dimension is $(N-2)^2+1$.
The coset space $SU(N)/(SU(N-2)\times U(1)\times U(1))$ has dimension $4N-6$, which
gives the dimensionality of the charge-breaking $SU(N)$-{\em orbit}, where the chosen
point lies. Since we have a one-parametric family of such orbits by varying $v^2_1$
and $v^2_2$ but keeping the sum ($r^0$) constant, we conclude that the dimension of
the charge-breaking $\vec n$-orbit space is $4N-5$. In the $r^\mu$-space, the
dimension is $4N-4$, which again coincides with the counting of
Section~\ref{section:fields}.

Now let us take a closer look at a point lying in the ``maximally charge-breaking''
orbit space, that is, the one with $K$-matrix conjugate to diag$(v^2,v^2,0,\dots,0)$.
It corresponds to maximally charge breaking (MCB) orbit because $s_2$,
which quantifies charge breaking, is maximum for a fixed $r_0$. It also corresponds
to vectors $\vec n$ with smallest $|\vec n|^2$ possible and, therefore, lying
on the surface of the inner cone.
Such a point has a larger isotropy group than a generic charge-breaking point.
Indeed, its isotropy group is now $SU(N-2)\times SU(2)\times U(1)$ of dimension
$(N-2)^2+3$, therefore the dimension of the coset space (and of the maximally
charge-breaking $\vec n$-orbit space) is $4N-8$. When considering $r^\mu$, it
corresponds to a manifold of dimension $4N-7$.

To summarize, we can group the $SU(N)$-orbits into classes of orbits according to its
isotropy groups. A set of orbits with the same isotropy group is called a
\textit{stratum}\,\cite{MR}. For our problem, we have in general three strata for a
fixed $r_0$:
\begin{itemize}
\item[(I)\hs{1ex}]
$\vec{n}^2=1$, one (neutral) orbit, isotropy group $SU(N-1)\otimes U(1)$.
\item[(II)\hs{.5ex}]
$\vec{n}^2 \in (a^2_N,1)$, continuous set of (charge-breaking)
orbits, isotropy group $SU(N-2)\otimes U(1)\otimes U(1)$.
\item[(III)]
$\vec{n}^2 =a^2_N$,
one (maximally charge-breaking) orbit, isotropy group
$SU(N-2)\otimes SU(2)\otimes U(1)$. \end{itemize}
Notice that for $N=3$ the strata I and III have the same isotropy group.

As a final remark, we note that $SU(N)/SU(N-1)$, in fact, defines the space of
$N$-complex-dimensional vectors of unit absolute value (i.e. sphere $S^{2N-1}$). Its
coset space with respect to the group $U(1)$ of the overall phase rotations,
$(SU(N)/SU(N-1))/ U(1) = SU(N)/(SU(N-1)\times U(1))$, is by definition the complex
projective space $\mathbb{CP}^{N-1}$. Thus, we recover the shape of the neutral orbit
space just from its isotropy group.

\section{Orbit space of 3HDM}
\label{sec:3hdm}
\subsection{The three sets of coordinates}

In this Section we analyze the orbit space of the three-Higgs-doublet model in more detail.

The Higgs field space of the 3HDM has 12 dimensions, hence the dimensionality of
the charge-breaking and neutral orbit spaces are 8 and 5, respectively. They are embedded in the 9-dimensional space
of $(r_0, r_i)$, $i = 1,\dots,8$, and are limited to the interior and the surface of the forward cone $r_0^2 - \vec r^2 = 0$.
In the 8D space of ``normalized'' vectors $\vec n \equiv \vec r/r_0$, 
the charge-breaking and neutral orbit spaces are 7D and 4D, respectively.

The $K$-matrix is a hermitean $3\times 3$ matrix, which is decomposed via the unit
matrix and the Gell-Mann matrices $\lambda_i$,
$i=1,\dots, 8$:
\be
\label{K:3HDM}
K = r_0 \cdot {1\over\sqrt{3}}{\bf 1}_3 + r_i \lambda_i\,.
\ee
The explicit expressions for the coordinates are
\bea
&& r_0 = {1 \over\sqrt{3}} \Tr K = {1 \over\sqrt{3}} \left(\phi_1^\dagger\phi_1 + \phi_2^\dagger\phi_2 + \phi_3^\dagger\phi_3\right)\,.\\
&& 
r_i = {1\over 2}\Tr[K \lambda_i]\,, \quad r_3 = {(\phi_1^\dagger\phi_1) - (\phi_2^\dagger\phi_2) \over 2}\,,\quad
r_8 = {(\phi_1^\dagger\phi_1) + (\phi_2^\dagger\phi_2) - 2(\phi_3^\dagger\phi_3) \over 2\sqrt{3}}\,,\quad
\nonumber\\
&&r_1 = \Re(\phi_1^\dagger\phi_2)\,,\ r_2 = \Im(\phi_1^\dagger\phi_2)\,,\
r_4 = \Re(\phi_1^\dagger\phi_3)\,,\nonumber\\ &&r_5 = \Im(\phi_1^\dagger\phi_3)\,,\
r_6 = \Re(\phi_2^\dagger\phi_3)\,,\ r_7 = \Im(\phi_2^\dagger\phi_3)\,.
\eea
It is also useful to group the last six real coordinates (which we will refer to as the ``transverse'' coordinates) 
into three ``complex coordinates'':
\be
r_{12} = r_1 + i r_2\,,\quad r_{45} = r_4 - i r_5\,,\quad r_{67} = r_6 + i r_7\,.\label{complexcoordinates}
\ee 
The same indices accompany the normalized vectors $\vec n$. 

\begin{figure}[!htb]
\centering
 \includegraphics[height=6cm]{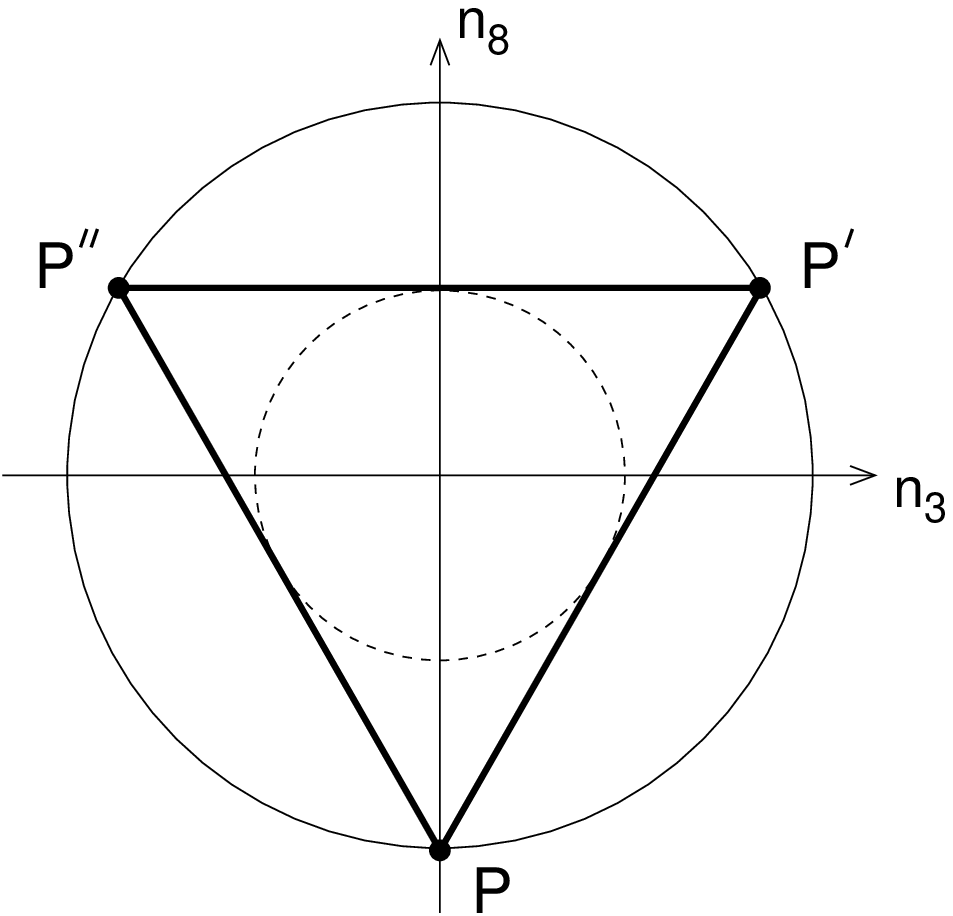}\hspace{3cm}
 \includegraphics[height=6cm]{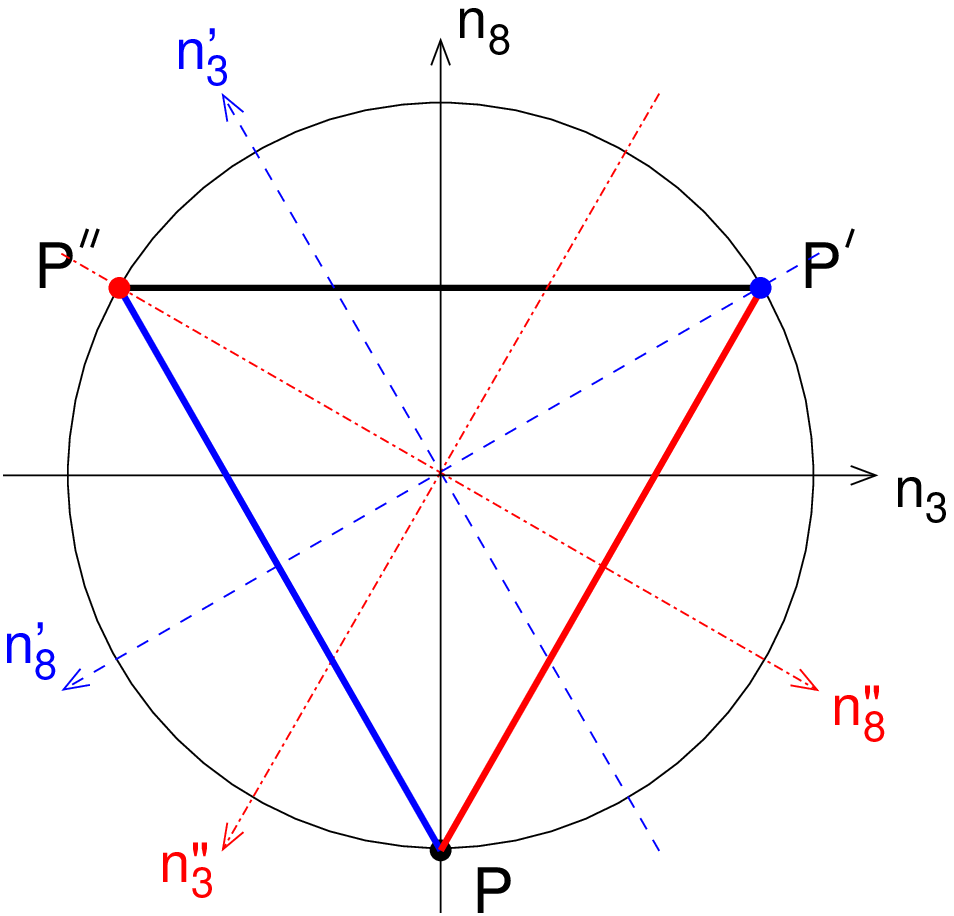}
\caption{(Left) The $(n_3,n_8)$-plane; all other $n_i=0$. Shown are the unit circle (section of the lightcone),
the inner circle (dashed line), the three points $P$, $P'$, $P''$, from the neutral manifold, and the three line segments
from the charge-breaking manifold. (Right) The same plane, but with three sets of coordinates: $(n_3,n_8)$ shown in black solid lines, 
$(n'_3,n'_8)$ shown in dashed blue lines, and $(n''_3,n''_8)$ shown in dash-dotted red lines.}
\label{fig-n3n8plane}
\end{figure}

The root space of the 3HDM is represented by the $(n_3,n_8)$-plane (all the other $n_i = 0$), 
shown in Fig.~\ref{fig-n3n8plane}, left.
The neutral manifold intersects this plane by three distinct points $P$, $P'$, $P''$:
\bea
P: & K \propto \mathrm{diag}(0,0,1)\,, & n_3 = 0,\, n_8 = -1\,;\nonumber\\
P': & K \propto \mathrm{diag}(1,0,0)\,, & n_3 = {\sqrt{3}\over 2},\, n_8 = {1 \over 2}\,;\label{pointsPPP}\\
P'': & K \propto \mathrm{diag}(0,1,0)\,, & n_3 = -{\sqrt{3}\over 2},\, n_8 = {1 \over 2}\,.\nonumber
\eea
The charge-breaking manifold is represented by the three line segments joining these three points.
Thus, the full orbit space in the root plane is given by the equilateral triangle (the 2-simplex).
Note that this triangle lies in the annular region between the circles
of radii $1/2$ and 1, in compliance with (\ref{nmin}). 

The triangle has the $S_3$ symmetry, however the choice of coordinate used to describe it, $n_3$ and $n_8$, breaks it.
To restore this symmetry in the description, we introduce two additional coordinate sets on the same plane: 
$(n'_3,n'_8)$ and $(n''_3,n''_8)$, which are shown in Fig.~\ref{fig-n3n8plane}, right, by blue dashed and red dash-dotted axes.
These coordinate sets are obtained from $(n_3,n_8)$ by $2\pi/3$ and $4\pi/3$ rotations, respectively:
\be
n'_3,\, n''_3 = -{1\over 2} n_3 \pm {\sqrt{3} \over 2} n_8\,,\quad
n'_8,\, n''_8 = \mp {\sqrt{3}\over 2} n_3 - {1 \over 2} n_8\,.
\ee
Each of the points $P$, $P'$, $P''$ can be associated with its ``natural'' coordinate set:
\be
P:\ n_3 = 0,\, n_8 = -1\,,\quad
P':\ n'_3 = 0,\, n'_8 = -1\,,\quad
P'':\ n''_3 = 0,\, n''_8 = -1\,.\label{pointsPPP2}
\ee
The coordinates $n_8$, $n_8'$, $n_8''$ are closely related to the three barycentric
coordinates
\be
p ={1-2n_8 \over 3}\,,\quad p'={1-2n_8' \over 3}\,,\quad p''={1-2n_8'' \over 3}\,,\quad p+p'+p''=1\,,
\label{barycentric}
\ee
which are proportional to the distances from a given point on the root plane to the three edges of the triangle.
The three edges of the triangle, which describe the charge-breaking orbit space on the root plane, 
can be naturally parametrized by $p=0$, $p'=0$, and $p''=0$.

Thus, the points on the root plane can be described in a symmetric fashion using
either $\{n_3,n'_3,n''_3\}$ with $n_3+n'_3+n''_3=0$, or $\{n_8,n'_8,n''_8\}$ with
$n_8+n'_8+n''_8=0$, or $\{p,p',p''\}$ with $p+p'+p''=1$.

This symmetric description can be extended to the entire orbit space $\calV_\Phi$.
Indeed, the $2\pi k/3$-rotations on the root plane are generated by a cyclic permutation of doublets:
\be
\{\phi'_1,\phi'_2,\phi'_3\} = \{\phi_2,\phi_3,\phi_1\}\,,\quad
\{\phi''_1,\phi''_2,\phi''_3\} = \{\phi_3,\phi_1,\phi_2\}\,.
\ee
This permutation changes the ``transverse'' coordinates (\ref{complexcoordinates}) according to
\be
\{n'_{12},n'_{45},n'_{67}\} = \{n_{67},n_{12},n_{45}\}\,,\quad
\{n''_{12},n''_{45},n''_{67}\} = \{n_{45},n_{67},n_{12}\}\,.
\ee
In other words, structures in the entire orbit space can be described in an explicitly $S_3$-symmetric way 
using coordinates
\bea
\label{symmetricset}
p,\ p',\ p'',\ n_{12},\ n'_{12},\ n''_{12}\,,\quad \mbox{where}\quad p+p'+p''=1\,.
\eea
The unit sphere is represented in terms of these coordinates as
\be
{3(p^2 + p^{\prime 2} + p^{\prime \prime 2})-1 \over 2} + |n_{12}|^2 + |n'_{12}|^2 + |n''_{12}|^2 = 1\,.
\ee
The $K$-matrix of 3HDM also takes a very symmetric form:
\be
K_{ab} \equiv \phi^\dagger_b\phi_a = r_0 \left(
\begin{array}{ccc} 
\sqrt{3}p' &  n_{12}^* & n_{12}'' \\[2mm]
n_{12} & \sqrt{3}p'' & n_{12}^{\prime *} \\[2mm]
n_{12}^{\prime\prime *} &  n_{12}' & \sqrt{3}p
\end{array}
\right)\,.\label{Kmatrixsymm}
\ee
Finally, the three quantities $z_{ab}$ (\ref{s2:zab}) can be written as
\be
z_{12} = 3 p' p'' - |n_{12}|^2 \ge 0\,,\quad
z_{13} = 3 p p' - |n_{12}^{\prime\prime}|^2 \ge 0\,,\quad
z_{23} = 3 p'' p - |n_{12}^\prime|^2 \ge 0\,.\label{zab:3HDM}
\ee

\subsection{$d$-condition}

In 3HDM, the $K$-matrix is a positive-semidefinite matrix with zero determinant, \cite{heidelberg}; 
thus, the list of constraints on the coordinates of $r^\mu$
truncates\,\footnote{One can check explicitly
that the higher order equations become identities. 
For example, thanks to the relation $d_{ijc}d_{ckl}+d_{jkc}d_{cil}+d_{kic}d_{cjl}=
(\delta_{ij}\delta_{kl}+\delta_{jk}\delta_{il}+\delta_{ki}\delta_{jl})/3$,
which holds for $N=3$, we get 
$\Gamma^{(4)}_{ijkl}r_ir_jr_kr_l = \vr^4$, which makes $s_4(K) =0$ satisfied
automatically.} 
at (\ref{s:3}),
which we will refer to as the ``$d$-condition''.
In the $\vec n$-space this condition can be written as
\be
\label{dcondition}
\sqrt{3}d_{ijk} n_i n_j n_k = {3 \vec n^2 - 1\over 2}\,.
\ee
In order to select out the neutral manifold, we accompany the $d$-condition with $\vec n^2 = 1$, 
which makes it
\be
\label{dcondition:neutral}
\sqrt{3}d_{ijk} n_i n_j n_k = 1\,.
\ee
Alternatively, the neutral manifold can be defined even more compactly 
with the aid of the ``star-product'' $(\vec m*\vec n)_k \equiv \sqrt{3}d_{ijk} m_i
n_j$ ($\vee$-product in Ref.\,\cite{nishi2006}):
\be
\vec n^2 = 1\,,\quad \vec n*\vec n = \vec n\,.
\ee
Let us now write the $d$-condition (\ref{dcondition}) explicitly using the well-known values of $d_{ijk}$:
\bea
&&3\cdot {\sqrt{3} \over 2}n_3 (n_4^2+n_5^2 - n_6^2-n_7^2) - n_8^3 + 
3\cdot n_8\left(n_1^2+n_2^2+n_3^2 - {n_4^2+n_5^2+n_6^2+n_7^2\over 2}\right)\nonumber\\
&& + 6\cdot {\sqrt{3} \over 2}(n_1n_4n_6 + n_1n_5n_7 - n_2n_4n_7 + n_2n_5n_6) = {3\vec n^2-1 \over 2}\,.
\eea
It can be rewritten in terms of symmetric coordinates (\ref{symmetricset}):
\be
p|n_{12}|^2 + p'|n'_{12}|^2 + p'' |n''_{12}|^2 - 3 p p' p'' - {2 \over \sqrt{3}}\Re(n_{12}n'_{12}n''_{12}) = 0\,,
\label{shape3}
\ee
which exposes the $S_3$ symmetry of the orbit space.
One can also arrive at this expression directly from det$K=0$ using representation (\ref{Kmatrixsymm}) for the $K$-matrix.

For the neutral manifold, the $d$-condition can be simplified further.
Let us recall that the lightcone condition (\ref{LCcondition}) implies that all three $z_{12}$, $z_{23}$, $z_{31}$ are equal to zero.
Using (\ref{zab:3HDM}), and denoting the sum of the phases of $n_{12}, n'_{12}, n''_{12}$ as $\gamma$,
one can cast the $d$-condition for the neutral orbit space into
\be
p p' p''(1-\cos\gamma) = 0\,.\label{shape5}
\ee

\subsection{The local properties of the orbit space}

So far, we have described the shape of the orbit space on the root plane $(n_3, n_8)$, 
with all the transverse coordinates $n_{12} = n_{45} = n_{67} = 0$.
Let us now gain an intuitive picture of how the orbit space extends into the
transverse space.

\begin{figure}[!htb]
   \centering
 \includegraphics[height=5cm]{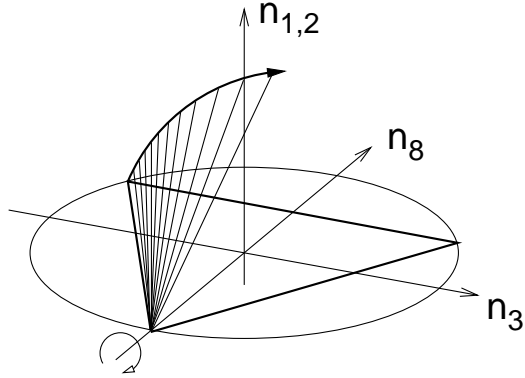}
\caption{Generating a 4D cone from the triangle by mixing the first two doublets.}
   \label{fig-rotationsn123}
\end{figure} 

In principle, the entire orbit space can be reconstructed by applying the full group
of adj$SU(3)$ of orthogonal transformations of $\vec n$ induced by unitary $SU(3)$
transformations among the doublets to the triangle on the root plane.
To make this result more visual, let us first consider the subgroup of adj$SU(3)$
induced by $SU(2)$ transformations between the first two doublets:
\be
\phi_a \to \bar\phi_a = R_{ab} \phi_b\,,\quad 
R_{ab} = 
\left(\begin{array}{ccc} 
\cos\ml{\!\frac{\alpha}{2}}\,e^{i\gamma}
& \sin\ml{\!\frac{\alpha}{2}}\,e^{-i\beta} & 0 \\[2mm]
-\sin\ml{\!\frac{\alpha}{2}}\,e^{i\beta}
& \cos\ml{\!\frac{\alpha}{2}}\,e^{-i\gamma} & 0 \\[2mm]
0 & 0 & 1 \end{array}\right)\,. 
\ee
The corresponding transformation of vectors $n_i \to \bar n_i$ brings a point on the root plane to the point with coordinates
\bea
&&\bar n_1 = - \sin\alpha\,\cos(\beta-\gamma)\, n_3\,,\quad
\bar n_2 = - \sin\alpha\,\sin(\beta-\gamma)\, n_3\,,\nonumber\\
&& \bar n_3 = \cos\alpha\, n_3\,,\quad
\bar n_8 = n_8\,,\quad \bar n_{45} = \bar n_{67} = 0\,.
\eea
The $SU(2)$ subgroup of such transformations, which we call ${\cal R}$-rotations, 
applied to the triangle sends it
to the surface of a 4D cone lying in the $n_{45}=n_{67}=0$ subspace
with the apex at point $P$, which is schematically illustrated by Fig.~\ref{fig-rotationsn123}. 
Indeed, the upper edge of the triangle is mapped to the 3D ball
\be
n_1^2 + n_2^2 + n_3^2 \le {3 \over 4}\,,\quad n_8 = {1 \over 2}\,,\quad n_{45}=n_{67} = 0\,,
\ee
which serves as the base of the cone and which is nothing else but the orbit space of the 2HDM.
The two other edges of the triangle are mapped to the lateral surface of the cone
\be
n_1^2 + n_2^2 + n_3^2 = {(1+n_8)^2 \over 3}\,.
\ee
Note that the neutral orbit space is represented in this 4D cone by the apex 
and by the ``rim'' of the based, the sphere $n_1^2 + n_2^2 + n_3^2 = 1$ at $n_8=1/2$.
 
The similar constructions arise from mixing of other pairs of the doublets.
Namely, the $SU(2)$ subgroup that mixes $\phi_2$ and $\phi_3$ (${\cal R}'$-rotations) keeps point $P'$ invariant
and sends the triangle to a 4D cone with the base
\be
n_1^{\prime 2} + n_2^{\prime 2} + n_3^{\prime 2} \le {3 \over 4}\,,\quad n'_8 = {1 \over 2}\,,
\ee
which lies in the subspace $n_{12}=n_{45} = 0$. Finally, mixing $\phi_1$ and $\phi_3$ (${\cal R}''$-rotations) generates
a similar cone with apex at $P''$ lying in the subspace $n_{12}=n_{67} = 0$.
 
Thus, in very loose terms, the shape of the orbit space can be described as follows: 
it is a manifold stretched between differently oriented three 4D cones. However, when visualizing 
this picture, one should remember that in fact there is no distinction between the base, the lateral surfaces
of these cones and of the part of the orbit space that is stretched between the cones.
The neutral orbit space, being nothing but $\mathbb{CP}^2$, 
is a {\em homogeneous} manifold, so it looks the same at every point. 
For the charge-breaking orbit space, rather similarly, there is a ``flat face'' going through each point.
Some additional hints for visualization of $\mathbb{CP}^2$ are given in \cite{CPn,book-geometry}. 

To make these observations more precise, let us calculate the sectional curvatures along
all mutually orthogonal directions at any point in the orbit space.

We start with a point located on a charge-breaking manifold. By an appropriate
reparametrization transformation we bring it to the root plane and place it, for
example, on the upper side of the triangle, where its position is described by $n_8 =
1/2$ and some $n_3$. 
We know that this point lies inside a flat 3D ball. Hence, there are three directions
(parallel to axes $n_1$, $n_2$ and $n_3$), along which the orbit space is flat in the
vicinity of the selected point.

We are left with four other directions, along $n_4$, $n_5$, $n_6$, and $n_7$.
One can shift into these directions by performing small rotations introduced above.
Note that from the point of view of the ${\cal R}'$- and ${\cal R}''$-rotations, 
the upper edge of the triangle is located at the lateral edge
of the corresponding cone, which brings in some curvature.

Explicitly, let us apply to a point on the upper edge the sequence of 
an ${\cal R}'$-rotation with an infinitesimal $\alpha'$
and an ${\cal R}''$-rotation with an infinitesimal $\alpha''$, all the other angles
$\beta'$, $\gamma'$, $\beta''$, $\gamma''$ being arbitrary 
(the order of the two transformations is irrelevant for this calculation). 
We get shifts of $n_{45}$ and $n_{67}$ linear in the small angles 
\be
n_{67} \approx - {\sqrt{3} \over 4}\alpha' e^{i\beta'}\left(1-{2 \over\sqrt{3}}n_3\right)\,,
\quad
n_{45} \approx {\sqrt{3} \over 4}\alpha'' e^{i\beta''}\left(1+{2 \over\sqrt{3}}n_3\right)\,,
\ee
and shifts in $n_3$, $n_8$ which are quadratic in small angles:
\bea
\delta n_8 &\approx & - {3 \over 16}\left[\alpha^{\prime 2} \left(1-{2 \over\sqrt{3}}n_3\right) 
+ \alpha^{\prime\prime 2} \left(1+{2 \over\sqrt{3}}n_3\right) \right]\,,
\nonumber\\[2mm]
\delta n_3 &\approx& {\sqrt{3} \over 16}\left[\alpha^{\prime 2} \left(1-{2 \over\sqrt{3}}n_3\right) 
- \alpha^{\prime\prime 2} \left(1+{2 \over\sqrt{3}}n_3\right) \right]\,.
\eea
For the charge-breaking manifold, where $n_3$ is a flat direction, we need to keep track 
only of the changes in $n_8$, for which we get:
\be
\delta n_8 \approx - {|n_{67}|^2 \over 2 R_{67}} - {|n_{45}|^2 \over 2 R_{4,5}}\,,
\ee
where the curvature radii along directions $n_6$, $n_7$ ($R_{6,7}$) and along
directions $n_4$, $n_5$ ($R_{4,5}$), are
\be
R_{4,5} = {1 \over 2}\left(1+ {2\over\sqrt{3}}n_3\right)\,,\quad
R_{6,7} = {1 \over 2}\left(1- {2\over\sqrt{3}}n_3\right)\,.
\ee
Thus, the charge-breaking orbit space has locally the shape of an ellipsoidal
cylinder, with three flat directions and two pairs of curved directions with
sectional curvature radii $R_{4,5}$ and $R_{6,7}$.

We now repeat this calculation for a point at the neutral manifold, for example, point $P$. 
We again perform two infinitesimal rotations ${\cal R}'$ and ${\cal R}''$ and calculate shifts 
of the coordinates.
This time we must take care of shifts of all eight coordinates $n_i$.
These rotations give linear shifts in small $\alpha'$ and $\alpha''$ to the four transverse coordinates,
\be
\delta n_4 \approx {\sqrt{3}\over 2} \alpha''\cos\beta'' \,,\quad
\delta n_5 \approx {\sqrt{3}\over 2} \alpha''\sin\beta'' \,,\quad
\delta n_6 \approx {\sqrt{3}\over 2} \alpha'\cos\beta' \,,\quad
\delta n_7 \approx {\sqrt{3}\over 2} \alpha'\sin\beta' \,,
\ee
and quadratic shifts to the other coordinates
\be
\delta n_{12} \approx {\sqrt{3}\over 4}\alpha'\alpha''\, e^{i(\beta''-\beta')}\,,\quad
\delta n_3 \approx {\sqrt{3}\over 8}(\alpha^{\prime\prime 2} - \alpha^{\prime 2})\,,\quad
\delta n_8 \approx {3\over 8}(\alpha^{\prime 2}+ \alpha^{\prime\prime 2})\,.
\ee
The overall quadratic shift is
\be
\delta n = \sqrt{(\delta n_1)^2 + (\delta n_2)^2 + (\delta n_3)^2 + (\delta n_8)^2} \approx 
{\sqrt{3}\over 4} (\alpha^{\prime 2}+ \alpha^{\prime\prime 2}) \approx {|\delta n_{45}|^2 + |\delta n_{67}|^2 \over \sqrt{3}}\,.
\ee
Thus, the curvature radius of the neutral manifold is $R_0=\sqrt{3}/2$ regardless of the direction of the shift. 
Since this holds true at every point of the neutral manifold, it means that the neutral manifold 
is an example of spherical space forms (a manifold of constant sectional curvature).
This comes as no surprise: it is known that an even-dimensional spherical space must
be a sphere or a complex projective space\,\cite{book-spaces}.

Note that the curvature radius $R_0$ does not and should not coincide with the largest curvature radius 
of the charge-breaking manifold near the rim. 
In loose terms, the fact that the neutral orbit space is located on the 
unit sphere gives to the neutral points more curvature with respect to the adjacent charge-breaking points. 

\subsection{Duality property of the orbit space}

The orbit space of 3HDM has an additional {\em duality} property, which does not hold for a generic $N$:
if a ray along direction $\vec n$ goes through the neutral orbit space,
then a ray in the opposite direction, $-\vec n$, points towards a maximally charge-breaking point.
Since the charge-breaking points lie on the sphere $|\vec n|=1/2$, we find that 
the maximally charge-breaking orbit space is homothetic to the neutral orbit space with the scale factor of $1/2$.

This property can be easily proved in the root plane:
if the neutral point is characterized by the $K$-matrix
diag$(0,0,v^2)=\frac{v^2}{3}(\id-\sqrt{3}\lambda_8)$ ($|\vec n|=1$), then
the opposite point corresponds to the $K$-matrix
diag$(v^2,v^2,0)=\frac{2v^2}{3}(\id+\frac{\sqrt{3}}{2}\lambda_8)$
($|\vec n|=1/2$).
This property clearly depends on the number of the diagonal elements and does not
generalize for higher $N$.
However, at $N=4$ another observation can be made: if $\vec n$ points towards a
maximally charge-breaking point, then so does $-\vec n$. For
example $K=\diag(1,1,0,0)$ is opposite to $K=\diag(0,0,1,1)$ with the same $r_0$.
That is, the maximally charge-breaking orbit space in the four-Higgs-doublet model is
centrally symmetric.

\section{Conclusion}

In this paper we initiated an analysis of the general $N$-Higgs-doublet model.
Focusing only on the scalar sector of the model, 
we considered here a specific question: how to efficiently describe the space
of gauge-invariant bilinears of Higgs fields in NHDM (the orbit space). We characterized the orbit space
as a certain algebraic manifold embedded in the Euclidean space $\RR^{N^2}$ and
studied some of its algebraic and geometric properties. The general construction was
illustrated with the case of $N=3$, for which more detailed calculations were
presented.

For general NHDMs for $N>2$, compared to the $N=2$ case, there arises a general and
distinct feature of the orbit space:
the orbit space is no longer convex, i.e., for two arbitrary points
$x^\mu$ ($K_1$) and $y^\mu$ ($K_2$) in $\calV_\Phi$ [$M_h^*(N;2)$], the line segment
joining them may not be entirely contained in $\calV_\Phi$
[$M_h^*(N;2)$]\,\cite{nishi:nhdm}.
For example, for $K_1=\diag(v^2,v^2,0)$ and $K_2=\diag(0,u^2,u^2)$, their middle
point is $\frac{1}{2}(K_1+K_2)=\frac{1}{2}\diag(v^2,v^2+u^2,u^2)$ which no longer has
rank 2 or smaller.
The exception is the case of two neutral points, as explained in
Sec.\,\ref{subsec:geometric}.
More particularly, we showed there is a ``hole'' in the orbit space of constant
$r_0$, such that in $\vec n$-space it is constrained inside the
annular region of radius $|\vec n|=1$ (lightcone) and $|\vec n|=a_N$ (inner cone).
In other words, for $r_0>0$, we can not reach $|\vr|< a_Nr_0$. This feature will
bring very distinct possibilities to the symmetry breaking patterns of the potential
as well as to the conditions for bounded below potentials.
Some of its consequences will be further detailed in a forthcoming
work\,\cite{ivanov2010}.

We also commented on a remarkable similarity between the orbit space of NHDM and
the state space of an $N$-qudit in quantum information theory. We sketched a small ``dictionary''
between some objects in these two branches of theoretical physics, 
and we think that this link should be explored further.

The next step of this analysis, the study of the NHDM Higgs potential and its
symmetries, is done in the companion paper \cite{ivanov2010}. That study is also
conducted in the orbit space and uses many of the results of the present paper.
We hope that the methods presented in these papers will boost systematic exploration
of the wealth of structures hidden in the general NHDM.

It is clear that a very similar mathematics arises not only in multi-doublet models,
but also in models with $N$ copies of Higgs fields in other representations (scalars, triplets, etc).
It is therefore conceivable that even more complicated Higgs sectors can be treated along these lines.
Other possible applications could be found in the condensed matter physics, 
where group-invariant potentials depending on several interacting order parameters
are often used, \cite{landaubook}. An example where the methods of 2HDM were used to understand the general Ginzburg-Landau model
with two order parameters can be found in \cite{GL}.
\\

\section*{Acknowledgements}
This work was initiated by very stimulating discussions
with P.~Ferreira and J.~P.~Silva. Useful discussions with J.-R.~Cudell are also
acknowledged.
The work of I.P.I. was supported by the Belgian Fund F.R.S.-FNRS via the
contract of Charg\'e de recherches and it part by grants
RFBR No.08-02-00334-a and NSh-3810.2010.2.
The work of C.C.N. was partially supported by the Brazilian Agencies FAPESP and CNPq
through the grants 09/11309-7 and 309455/2009-0, respectively.

\appendix

\section{Relations among $z_{ab}$}\label{appendix:zab}

In order to show that among $N(N-1)/2$ quantities $z_{ab}$ in the NHDM there are only $2N-3$ algebraically 
independent quantities, we need to prove the following statement.
Take any four doublets, e.g. $\phi_1$ to $\phi_4$ with known norms, $(\phi_a^\dagger \phi_a)$;
suppose that $z_{12}, z_{13}, z_{14}, z_{23}, z_{24}$ are also known (we assume here a generic situation
when all $z_{ab}$ are non-zero). Then $z_{34}$ is not independent and can take 
at most two different values. If all of $z_{12}, z_{13}, z_{14}, z_{23}, z_{24}$ happen to be zeros,
then $z_{34} =0$.

We first note that the doublets are vectors in the space $\mathbb{C}^2$.
Therefore, if $z_{12} \not = 0$, all the doublets can be decomposed in the basis of $\phi_1$ and $\phi_2$:
\be
\label{decomposition34}
\phi_3 = c_1 \phi_1 + c_2 \phi_2\,,\quad \phi_4 = d_1 \phi_1 + d_2 \phi_2\,.
\ee
Note that the absolute value of the scalar product $(\phi_1^\dagger \phi_2)$ is known,
$|(\phi_1^\dagger \phi_2)|^2 = (\phi_1^\dagger \phi_1)(\phi_2^\dagger \phi_2) - z_{12}$,
its phase $\theta_{12}$ is not.
Let us now introduce the ``vector'' product of two doublets:
\be
[\phi_a \times \phi_b] \equiv \phi_a^+\phi_b^0 - \phi_a^0 \phi_b^+\,,
\ee 
where superscripts $+$ and $0$ refer to the upper and lower components of the doublets.
One can check that 
\be
[\phi_a \times \phi_b]^*[\phi_a \times \phi_b] = 
(\phi_a^\dagger \phi_a)(\phi_b^\dagger \phi_b) - (\phi_a^\dagger \phi_b)(\phi_b^\dagger \phi_a) = z_{ab}\,.
\ee
This leads to
\be
|c_1|^2 = {z_{23} \over z_{12}}\,,\quad |c_2|^2 = {z_{13} \over z_{12}}\,,\quad 
|d_1|^2 = {z_{24} \over z_{12}}\,,\quad |d_2|^2 = {z_{14} \over z_{12}}\,.
\ee
Therefore, decomposition (\ref{decomposition34}) turns into
\be
\phi_3 = \sqrt{{z_{23} \over z_{12}}} e^{i\eta_1} \phi_1 + \sqrt{{z_{13} \over z_{12}}} e^{i\eta_2} \phi_2\,,\quad 
\phi_4 = \sqrt{{z_{24} \over z_{12}}} e^{i\xi_1} \phi_1 + \sqrt{{z_{14} \over z_{12}}} e^{i\xi_2} \phi_2\,.
\ee
The phase differences $\eta_2-\eta_1$ and $\xi_2-\xi_1$ are both related to the (unknown) phase $\theta_{12}$:
\bea
z_{12}(\phi^\dagger_3 \phi_3) &=& z_{23}(\phi^\dagger_1 \phi_1) + z_{13}(\phi^\dagger_2 \phi_2)
+ 2\sqrt{z_{13}z_{23}}|(\phi^\dagger_1 \phi_2)|\cos(\theta_{12}+\eta_2-\eta_1)\,,\nonumber\\
z_{12}(\phi^\dagger_4 \phi_4) &=& z_{24}(\phi^\dagger_1 \phi_1) + z_{14}(\phi^\dagger_2 \phi_2)
+ 2\sqrt{z_{14}z_{24}}|(\phi^\dagger_1 \phi_2)|\cos(\theta_{12}+\xi_2-\xi_1)\,.
\eea 
Now, the quantity $z_{34}$ can be written as
\be
z_{12} z_{34} = z_{23}z_{14} + z_{24}z_{13} - 2\sqrt{z_{23}z_{14}z_{24}z_{13}}\cos(\eta_1 + \xi_2 - \eta_2 - \xi_1)\,.
\ee
But 
\bea
&&\cos(\eta_1 + \xi_2 - \eta_2 - \xi_1) = \cos[(\theta_{12} + \xi_2 - \xi_1) - (\theta_{12} + \eta_2 - \eta_1)]\nonumber\\
&&\cos(\theta_{12} + \xi_2 - \xi_1)\cos(\theta_{12} + \eta_2 - \eta_1)
\pm |\sin(\theta_{12} + \xi_2 - \xi_1)\sin(\theta_{12} + \eta_2 - \eta_1)|\,,\nonumber
\eea
which can be expressed in terms of known cosines of $\theta_{12} + \xi_2 - \xi_1$ and $\theta_{12} + \eta_2 - \eta_1$.
This proves an algebraic relation between $z_{34}$ and the other quantities without the need to know $\theta_{12}$.
The sign ambiguity here means that two different values of $z_{34}$ can result.
However, if $z_{23}z_{14}z_{24}z_{13} = 0$, then $z_{34}$ is uniquely determined.

Note that if all of $z_{12}, z_{13}, z_{14}, z_{23}, z_{24}$ happen to be zero, it means that all four doublets are proportional
to each other, and therefore, $z_{34}$ must be zero as well.

As a remark, let us analyze in more generality the phenomenon of multidimensional
reduction imposed by a single condition\,\eqref{LCcondition}.
A more general situation can be envisaged.
The space of $N\times N$ hermitean matrices with rank equal or lower than
$r\le N$, which we can denote by $M_h(N;r)$, has dimension $r(2N-r)$.
To define $M_h(N;r)$, we need $s_n(K)=0$, $r\le n\le N$.
Despite only one constraint $s_{r-1}(K)=0$ being further required to restrict
$M_h(N;r)$ to $M_h(N;r-1)$, the dimensionality is indeed reduced by $2(N-r)+1$. In
our case, we have $r=2$ and the amount of dimensional reduction from $M_h(N;2)$ to
$M_h(N;1)$ is exactly $2N-3$. Therefore, any single condition $s_{r-1}(K)=0$
necessary to restrict $M_h(N;r)$ to $M_h(N;r-1)$ should contain multiple independent
conditions in the same way $s_2(K)=0$ is equivalent to various conditions
$z_{ab}=0$, as proved in this appendix.

\section{Maximal set of gauge invariants}
\label{ap:maximal}

We will show here how we can choose a maximal set of algebraically independent gauge
invariants
$\phi_b^\dag\phi_a=K_{ab}$, corresponding to the $4N-4$ degrees of freedom of the $N$
doublets $\phi_a$. If all bilinears $\phi_b^\dag\phi_a$, $a,b=1,\ldots,N$,
$a\le b$, were functionally independent, $N^2$ real parameters would be necessary for
parametrization.

Firstly, we should use the fact that a general non-null $K$ matrix\,\eqref{K:def} has
rank two or one. If it has rank two, it is always possible to choose a set of two
linearly independent lines (columns) of $K$ as a basis of the space spanned by all
the $N$ lines (columns); otherwise only one line is linearly independent and this
case can be treated easily. By appropriately labeling the doublets we can
choose the first and second lines to be non-null and non-parallel. In that case,
since $K$ is a hermitean matrix, we can choose the set in Eq.\,\eqref{minimal:K}
as the minimal set of gauge invariants.
It is easy to see that they can be parametrized by $4N-4$ real parameters,
considering that $K_{11},K_{22}$ are real.
We should assume $K_{11}\neq 0$ and $K_{22}\neq 0$ because, \textit{e.g.}, the case
$K_{11}=0$ directly implies $K_{1a}=K_{a1}=0$, $a>1$. Such property follows directly
from the fields language but it also can be thought as a consequence of
\eq{
\sum_{a\neq 1}|K_{1a}|^2 \le K_{11}\big(\sum_{a\neq 1}K_{aa}\big)\,,
}
which follows from the Schwarz inequality. Thus any null
diagonal element implies an entire null line and column of $K$.

It remains to be shown that all other $K_{ab}$, with $a,b\ge 3$, can be written
entirely in terms of the set in Eq.\,\eqref{minimal:K}.
Let us show how to calculate the elements in the third line. The calculation of any
other element follows analogously. By hypothesis, we can write any element in the
third line as a linear combination of the corresponding element in the first and
second lines:
\eq{
\label{K3a:alpha}
K_{3a}=\alpha K_{1a}+\beta K_{2a}\,,\quad a\ge 3\,.
}
But the the same coefficients $\alpha,\beta$ relate the elements in the first and
second columns as
\eqarr{
\label{K31}
K_{31}&=&K_{13}^*=\alpha K_{11}+\beta K_{21}\,,\cr
K_{32}&=&K_{23}^*=\alpha K_{12}+\beta K_{22}\,.
}
Equations \eqref{K31} can be rewritten as 
\eq{
\label{K31:matrix}
\begin{pmatrix}
K_{31} & K_{32}
\end{pmatrix}
=
\begin{pmatrix}
\alpha & \beta
\end{pmatrix}
K^{(2)}_{12}\,,
}
where $K^{(2)}_{ij}$ is a $2\times 2$ submatrix (minor) of $K$ containing only the
elements $K_{ab}$, with $a=i,j$ and $b=i,j$.
Thus we can invert equation \eqref{K31:matrix} to obtain
\eq{
\label{K3a:K12}
K_{3a}=
\begin{pmatrix}
K_{31} & K_{32}
\end{pmatrix}
(K^{(2)}_{12})^{-1}
\begin{pmatrix}
K_{1a} \cr	 K_{2a}
\end{pmatrix}
\,,
}
or
\eq{
\label{K3a:K12:2}
\det(K^{(2)}_{12})
K_{3a}=
\begin{pmatrix}
K_{31} & K_{32}
\end{pmatrix}
\mathrm{adj}(K^{(2)}_{12})
\begin{pmatrix}
K_{1a} \cr	 K_{2a}
\end{pmatrix}
\,,
}
where $\mathrm{adj}$ denotes the adjoint matrix. Notice $\det(K^{(2)}_{12})=z_{12}$
is non-null by hypothesis.

We can rewrite Eq.\,\eqref{K3a:K12} in a more compact form if we define the
2-dimensional complex vector
\eq{
\chi_{a}^{\tp}\equiv
\begin{pmatrix}
K_{1a} & K_{2a}
\end{pmatrix}
\,, \quad a=1,\ldots,N\,.
}
Then any matrix element $K_{ab}$ can be calculated as
\eq{
\label{Kab:K12}
K_{ba}=\chi_b^\dag(K^{(2)}_{12})^{-1}\chi_a\,.
}
Surprisingly, the expression in Eq.\,\eqref{Kab:K12} is valid not only for $a,b\ge
3$, but for all $a,b=1,\ldots,N$. However, for $a,b=1,2$, it leads to trivial
identities.

Equations \eqref{K3a:K12} and \eqref{Kab:K12} are direct consequences from the
fact that any $3\times 3$ submatrix of $K$ has null determinant for rank$K\le 2$.
For example, Eq.\,\eqref{K3a:K12:2} is equivalent to calculate the determinant of a
$3\times 3$ matrix constructed with the blocks
$K^{(2)}_{12},\chi_a,\chi_3^\dag,K_{3a}$ by cofactor expansion along the third
column.

One can identify the role of the coefficients $\alpha,\beta$ in
Eq.\,\eqref{K3a:alpha} if we recognize the equation as the expansion
\eq{
\phi_3=\alpha \phi_1+\beta\phi_2\,,
}
contracted to $\phi_a^\dag$. Hence, the coefficients of the linear expansion
\eq{
\phi_a=c_{a1} \phi_1+c_{a2}\beta\phi_2\,,
}
are solutions of
\eq{
\label{ca12}
\begin{pmatrix}
K_{a1} & K_{a2}
\end{pmatrix}
=
\begin{pmatrix}
c_{a1} & c_{a2}
\end{pmatrix}
K^{(2)}_{12}\,.
}
The hermitean conjugate of Eq.\,\eqref{ca12} can be also written
\eq{
\chi_a=c^*_{a1}\chi_1+c^*_{a2}\chi_2\,.
}

If rank$K$=1, we would have $K_{ba}=K_{b1}K_{1a}/K_{11}$, $K_{11}\neq 0$, for all
$a,b=1,\ldots,N$.

\section{Characterization of $SU(N)$ orbits}
\label{ap:invariants}

The vector space spanned by the $N\times N$ hermitean matrices, containing $K$, is
isomorphic to $\RR^{N^2}$, where the vectors $r^\mu$ live. The mapping between
these spaces were given by Eq.\,\eqref{K:decomposed} and it is valid even if
we generalize $K$ to be a general $N\times N$ hermitean matrix.
The action of the reparametrization group $SU(N)$ on $K$ is defined by
Eq.\,\eqref{K:U}.
Such action divides the space $\RR^{N^2}$ into $SU(N)$ orbits. Each of these orbits
can be uniquely characterized by a set of $N$ $SU(N)$ invariants functions $s_k(K)$,
$k=1,\ldots,N$, defined in Eq.\,\eqref{sn:def}.
Therefore, any orbit can be represented by a point in one connected region of a
$N$-dimensional diagram whose axes represent $s_k$. There is only one connected
region because we can vary the eigenvalues continuously, keeping, for instance, a
decreasing order.

The reparametrization group action in Eq.\,\eqref{K:U}, however, defines
naturally two invariant spaces (irreducible representations) which allows the
splitting 
\eq{
K=K_0+\tK\,,
}
where $K_0\equiv s_1(K)\id/N$ and $\tK=K-K_0=r_i\lambda_i$ is the traceless
part of $K$. Hence, $\tK$ is the component of $K$ that transforms non-trivially
under $SU(N)$ while $K_0$ is an invariant. This means that the invariants $s_k(K)$,
$k\ge 2$, are not fundamental but have contributions of the trivial part $K_0$,
already in $s_1(K)$.
We can use $s_k(\tK)$, instead of $s_k(K)$, for $k\ge 2$, which obviously are
invariant and does not depend on $K_0$. Let us denote $s_k\equiv s_k(K)$ and
$\ts_k\equiv s_k(\tK)$.
The relation between the two sets $\{s_k\}$ and $\{\tilde{s}_k\}$, $k=2,\ldots,N$,
can be obtained by comparing Eq.\,\eqref{characteristic:N} to
\eq{
\label{s:tilde}
\det(\lambda\id-K)=
\det(\tilde{\lambda}\id-\tK)=
\tilde{\lambda}^N+\sum_{k=2}^{N}(-1)^k \tilde{s}_k\tilde{\lambda}^{N-k}\,,
}
where $\tilde{\lambda}\equiv\lambda-\frac{s_1}{N}$.
The relation between $s_p$ and $\ts_p$, for $p\ge 3$, is
\eqarr{
\label{ts:s}
s_p-\ts_p&=&
-\sum_{k=1}^{p-2}
\ml{\binom{\!N+k-p}{k}}
\Big(\!\frac{\mfn{-}s_1}{\ms{N}	}\Big)^{k}s_{p-k}+
(p-1)
\ml{\binom{\!N}{\raisebox{.2em}{\mfn{p}}}}
\Big(\!\frac{\mfn{-}s_1}{\ms{N}}\Big)^{p}
\,,
\\&=&
\label{s:ts}
\sum_{k=1}^{p-2}
\ml{\binom{\!N+k-p}{k}}
\Big(\frac{s_1}{\ms{N}}\Big)^{k}\ts_{p-k}+
\ml{\binom{\!N}{\raisebox{.2em}{\mfn{p}}}}
\Big(\!\frac{s_1}{\ms{N}}\Big)^{p}
\,.
}
At last, all invariants $\ts_k=s_k\ms{(\tK)}$ can be calculated using
Eq.\,\eqref{sn:def} and written in terms of
\eq{
\label{tK:n}
\ml{\frac{1}{2}}\Tr[(r_i\lambda_i)^n]=
\Gamma^{(n)}_{i_1i_2\cdots i_n}r_{i_1}r_{i_2}\cdots r_{i_n}\,,
}
where the tensors $\Gamma^{(n)}$ were defined in Eq.\,\eqref{Gamma:n}.
For example, 
\eqarr{
\tilde{s}_2&=& -\ml{\frac{1}{2}}\Tr[(r_i\lambda_i)^2]=-\vr^2\,,\\
\label{st:2}
\tilde{s}_3&=&
\ml{\frac{1}{3}}\Tr[(r_i\lambda_i)^3]
=d_{ijk}r_ir_jr_k\,,
\\
\tilde{s}_4&=&
-\ml{\frac{1}{4}}\Tr[(r_i\lambda_i)^4+\ts_2(r_i\lambda_i)^2]
=
-\ml{\frac{1}{2}}\Gamma^{(4)}_{ijkl}r_{i}r_{j}r_kr_l
+\ml{\frac{1}{2}}\vr^4
\,.
}
All $\tilde{s}_k$ can be written in terms of the terms of Eq.\,\eqref{tK:n} with
equal or lower order.

It is important to notice that for general $N\times N$ hermitean matrices $K$, not
restricted to positive semidefinite rank two matrices, the characterization of the
$SU(N)$ orbits would involve more than one invariant, besides $r_0$. For instance,
for a value of $\vec{r}^2$, there would be infinitely many distinct orbits that have
to be further characterized by higher order invariants.

\end{document}